# Automated data-driven creation of the Digital Twin of a brownfield plant

Dominik Braun
Graduate School of Excellence advanced
Manufacturing Engineering (GSaME)
University of Stuttgart
Stuttgart, Germany
dominik.braun@gsame.uni-stuttgart.de

Wolfgang Schloegl
Digital Engineering
Siemens AG
Nuremberg, Germany
schloegl.wolfgang@siemens.com

Michael Weyrich
Institute of Industrial Automation and
Software Engineering
University of Stuttgart
Stuttgart, Germany
michael.weyrich@ias.uni-stuttgart.de

*Abstract*—The success of the reconfiguration of existing manufacturing systems, so called brownfield systems, heavily relies on the knowledge about the system. Reconfiguration can be planned, supported and simplified with the Digital Twin of the system providing this knowledge. However, digital models as the basis of a Digital Twin are usually missing for these plants. This article presents a data-driven approach to gain knowledge about a brownfield system to create the digital models of a Digital Twin and their relations. Finally, a proof of concept shows that process data and position data as data sources deliver the relations between the models of the Digital Twin.

I. INTRODUCTION

The production industry faces the challenge to deliver customer-specific products in decreasing time. Therefore, there is the need for an efficient reconfiguration procedure of existing production systems. This is called brownfield engineering and implies the reconfiguration and reuse of existing production systems, which are amended to new requirements, products or a technical progress. The reconfiguration has some limitations and challenges to reconfigure an existing system with minimal system downtime [1]. Development is already happening more frequently in a brownfield environment than in the greenfield because of the economic advantage [1, 2] and will become even more common in the future [3]. Greenfield engineering names the development of a completely new production system without limitations and specification from existing systems [1]. The Digital Twin as virtual representation of cyber-physical production systems provides the needed information and abilities to virtually create and test reconfiguration scenarios [4]. This simplifies the reconfiguration process and therefore reduces the downtime and errors due to system changes. The Digital Twin of brownfield systems is often missing or outdated because of previous modifications and small improvements that are usually poorly documented or not at all [3]. An adaptation of an existing documentation or, in the best case, a digital model is not carried out [5, 6] and thus cannot be consulted during the reconfiguration of a plant. The modification of brownfield plants, in particular the control software, is carried out directly in the plant during a reconfiguration using the trial-and-error method and is therefore one of the most time-consuming activities [5]. This results for instance in a higher reconfiguration duration, higher costs and a lower restart reliability [5]. But a high level of restart reliability after a reconfiguration is becoming increasingly important, especially due to the decreasing plant life cycles as a result of more variable product ranges [6] and the resulting need for more frequent reconfigurations [7]. Especially the control software of an industrial system which provides a large scope of the functions [8], is often reconfigured [9]. The reconfiguration process is a important but difficult task, due to various reasons. The authors in [10] summarize the most important factors such as high time-consumption or error-proneness which can interrupt the complete production. Thus, the reconfiguration needs to be supported for instance by a virtual commissioning. The Digital Twin provides such abilities to support the reconfiguration of a plant by virtual planning and commissioning [8]. Especially for reconfiguration processes, the Digital Twin can save up to 58% of the time in contrast to a manual planning and reconfiguration process [11]. A challenge is that for most existing plants, no digital models or complete, up-to-date documentations are available for the reasons mentioned at the beginning. Therefore, the digital models need to be created to use the Digital Twin to support the software reconfiguration.

*Structure*: Section II describe the state of the art in Digital Twins for manufacturing systems and how they can be created or synchronized. The following Section III presents the data driven approach, the used information sources and their analysis to create digital models to support the software reconfiguration. Section IV presents a proof of concept using a brownfield system to test the approach. Finally, Section V summarizes the results and provides an outlook about the future work.

*Objective*: This article presents a basics of a data-based approach to automatically create the multi-dimensional digital models of a Digital Twin to support the software reconfiguration of a production system. Therefore, the usage of position data is introduced to identify the physical dimensions of the underling mechatronic components of the PLC.

II. STATE OF THE ART

The components and abilities of a Digital Twin differ between the use cases and definitions in the literature. The authors in [4] analyzed the different definitions in literature and developed a common characterization by the intersection of the other definitions. These related, multi-dimensional models are the digital replica of the real asset and form the core of each Digital Twin (see Fig. 1). Besides that, the Digital Twin need to provide three characteristics additionally to the digital replica: some models are executable, the models are synchronized with the real asset and there need to be an active data acquisition from the real asset towards the Digital Twin. These abilities enable the digital replica to represent the static and dynamic behavior of the real asset over the entire lifecycle.

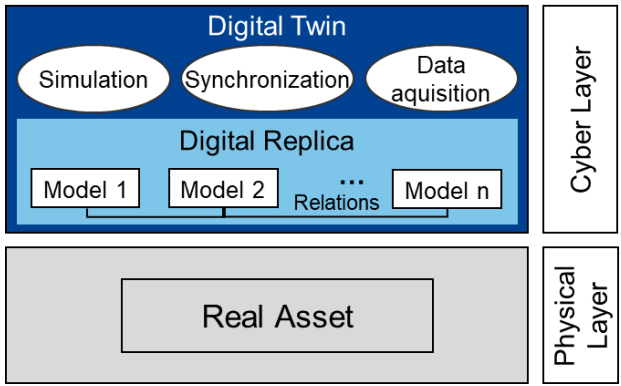

Fig. 1. Components and characteristics of a Digital Twin [4]

The models of a Digital Twin are usually created during the engineering phase of a production system in collaboration with many engineers and are integrated as Digital Twin using a product lifecycle management. These models and their relations are the fundamental base for the reconfiguration support as they enable the virtual commissioning of the existing plant. As stated in Section I these models are not synchronized with the plant or do not exist at all (in digital version) for brownfield production systems [5, 6]. The manual reengineering of the multi-dimensional digital models from the production system is very time consuming and error prone task. Therefore, automated methodologies are needed to create and synchronize the digital models of the current plant version. These approaches can be divided into cross-domain and domain specific approaches. They differed in the point whether a connection is made to the models of the other domains, mechanical, electrical or software.

*A. Cross-domain approach*

The most promising approach to automatically create cross-domain, up-to-date models and their relations is the anchor point method [12]. This method applies a naming convention following an industrial norm to all PLC elements (such as the I/O-signals or function blocks) that adds additional information about the related elements in the mechanical and electrical view of the Digital Twin. Because the PLC code of the running plant must be up-to-date, it can be used as source to update the Digital Twin. Changes to the real plant result in modifications in the PLC code. Comparing the current and an older version of the PLC code changes to the plant can be identified. Analyzing the defined names of the changed PLC elements, it is possible to automatically identify the related outdated item in the Digital Twin models. The affected elements in the Digital Twin then can be adapted to the plant. This method presumes a few requirements:

- The Digital Twin exists
- The PLC code related to the Digital Twin is available
- The PLC code after a modification is available
- The naming convention is considered in the PLC

For the reconfiguration of brownfield plants these requirements are not always fulfilled. The Digital Twin of existing plants is often missing and the needed naming convention for the anchor point method is not necessarily used.

*B. Domain-specific approach*

There are a few other methods available such as laser scans, photogrammetry or configuration analysis [13–16] to retrieve geometrical information about the real production system or the plant layout in a digital format. Other methods such the scan of paper-based redlined circuit documentations [17] address the model creation of electrical model or communication network scans to identify simulation behavior changes [18]. Each provides fundamental model information for the Digital Twin in a single domain but do not capture their relations towards others and the control software (elements). Furthermore, they are not providing information on the software and the information needed to reconfigure the control software of a production plant. These methods can be used to particularize the domain specific models of the Digital Twin after the base of the connected models is constructed.

III. DATA-DRIVEN DIGITAL TWIN CREATION APPROACH

The existing model creation and synchronization approaches only cover single domains of a Digital Twin or are not applicable to brownfield production systems to support the complicated reengineering task. Therefore, we propose an automated, data-driven approach to create the software-oriented models and their relation to the other dimensions of a Digital Twin. The models support the inspection, understanding and reconfiguration of the control software. With the relations to the other domains, engineers from other domains can easily reuse the created models, connect and enrich them with their domain specific information. It is also possible to combine this approach with other automated domain specific reengineering approaches such as the laser or document scanning (see Section II.B) because the created models already contain the cross-links towards the other domain models. Fig. 2 shows the scheme of the approach starting from the multiple information sources used representing the current plant status. In this research three data sources are used, the PLC, material position data and process data. Using different static and dynamic analysis approaches higher level knowledge can be extracted. The gained information by different analysis methods is integrated into a knowledge graph, for example in a labeled property graph. This knowledge representation performs well in relational analysis. Thus, it is used to connect and enrich the single information for the creation of a Digital Twin. The resulting knowledge graph is independent from the architecture used for the implementation and can be mapped into manufacturer-specific software systems to build a Digital Twin.

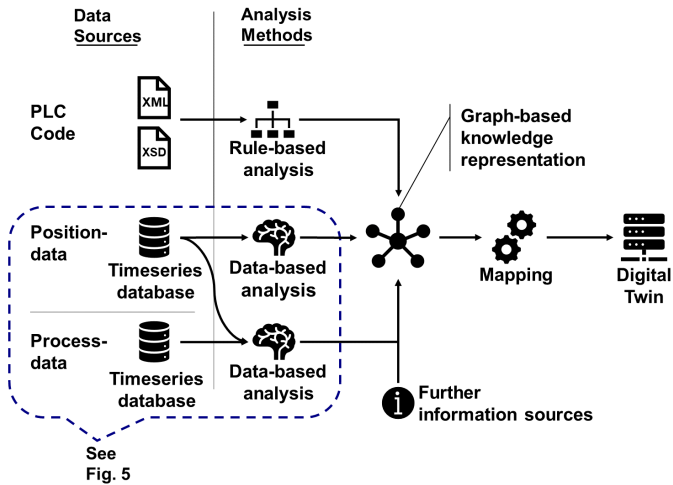

Fig. 2. Model creation approach using the data sources PLC (code and configuration), position data and process data from the real plant

There is no information source that provides all needed information about a production system to create a complete Digital Twin with all aspects. Even limited to the software perspective multiple information sources are needed to retrieve all the necessary information. The following sub-sections present the information sources used in this approach and their content. In Section IV.B the details of the analysis of position and process data are demonstrated using a brownfield testbed.

*A. Processing PLC code and configuration to identify mechatronic components*

With reference to the anchor point method, the PLC code and the PLC configuration are used as the starting point for the analysis. The PLC is a well-suited source to gain up-to-date information about the production system and its software because the PLC code is always updated otherwise the production fails. In contrast to the anchor point method, this approach does not focus on the name of the PLC elements but examines the contained elements and their dependencies. The elements such as function blocks, data blocks or hardware configurations are analyzed by fixed rules. To ensure these analyzes can be applied to different industrial plants and PLC vendors, the native PLC code is compiled into extensible markup language (XML) as manufacturer-independent exchange format. The most manufacturers provide these compilers together with the PLC development environment. The XML files can be parsed using an XML schema definition (XSD) file, that need to be created once for each vendor because the elements are structured slightly different between the producers. The imported PLC elements are analyzed by fixed rules from a decision tree and reproduce a model of the control software elements such as the signals or function blocks and their logical dependencies. These analyses reveal the structure and static information about the control software but it does not contain information to reconstruct the relation to other domains, the dynamic system behavior such as signal trends or the frequency of control sequences.

*B. Process data as information source of dynamical behavior*

Process data is all current and historical data that can be obtained from the plant control during operation. This includes sensor/actuator signals and plant parameters. Process data can be retrieved from a manufacturing plants using various communication technologies such as OPC UA or MQTT. These technologies are either directly supported by the PLC or can be upgraded by an additional module to the PLC. The data can then be tapped and stored in a database. Time series databases are suitable for storing these industrial signal traces. In order to be able to assign the dynamic signal traces to the associated signal in the PLC, the signal name must be stored in addition to the timestamp and the signal value. The process data is used to identify the dynamical behavior of the plant by analyzing the signal traces and their correlations to other signals.

*C. Location data providing the mechanical relations*

Another information source is location data of the manufacturing system resources and of the moving material inside the production system. There are several technologies available that are used to locate things, for instance autonomous mobile robots, inside production halls. Common setup used in this context are real time locating systems utilize ultra-wideband technology. Another technology that may be used in the future for localization in manufacturing plants is 5G. Besides the intended usage of these systems, the position data of the material flow can be used to determine the mechanical dimensions of the plant. The geometric information of the material positions about the plant is much less accurate than specifically developed technologies, such as laser scans, but also much faster and less expensive to retrieve. The rough accuracy of the geometric information from a real time locating system is sufficient to replicate the real plant in digital models to support the software reconfiguration and enable connecting to mechanic features. The extraction of the position data from the real time locating system depends on the system interface and must be adapted once for the different manufacturers. Some of these systems already have the ability to automatically store all data directly in a database. Once the position data has been stored in a database either directly or via an interface, it can be further processed independently of the hardware.

*D. Data-fusion of process and location data*

Additional value can be gained from the fused analysis of process data and position data as they connect the software domain with the mechanical domain. For this purpose, the data from the two sources are linked by their timestamps. The moving material is assumed as one major reason for changing sensor signals. Therefore, the database is filtered for changing sensor signals and position data is searched with matching timestamps. Using stochastic methods, the approximate positions of a sensor can be determined. The position of the sensor here corresponds to the switching position, the installation position can deviate from this for sensors. With reference for the software reconfiguration support, this switching position, which in the best case closely coincides with the real position, is sufficiently accurate and represents the relevant process event. This will be demonstrated in the next section using a real cyber-physical production system.

IV. PROOF OF CONCEPT

Using a cyber-physical production system, a proof of concept is presented which demonstrates that the data-driven approach delivers the information about the production system, which are needed to create the Digital Twin. The model structure of the mechatronic system and the relation between the single components are reconstructed as fundamental part of the Digital Twin. Therefore, a production system was designed and built using actual industrial machinery on the research campus ARENA2036 (Active Research Environment for the Next Generation of Automobile) to investigate and demonstrate the possibilities of flexible production systems and their Digital Twin (see Fig. 3). The flexible production system serves as brownfield testbed and is used for a proof of concept of the presented approach to create the Digital Twins models. The analysis of the data driven part of the concept using process and position data is tested using data from the real production system. The production system consists of three automated units (welding machine, movable robot and an intelligent warehouse) each have their own decentralized control unit to control and provide the respective services. An additional head control unit is responsible for coordinating the units to produce a model car from four sheet metal parts. The units are responsible for the following tasks: The intelligent warehouse provides prefabricated metal parts in work-piece carriers and withdraws empty carriers. The material flow inside this warehouse is controlled by 24 actuators and monitored by 37 sensors, which represent the current state of occupation. Additionally, a RFID system is installed to identify the workpiece carriers and their content. The warehouse PLC provides the necessary information to the other system participants via industrial WLAN.

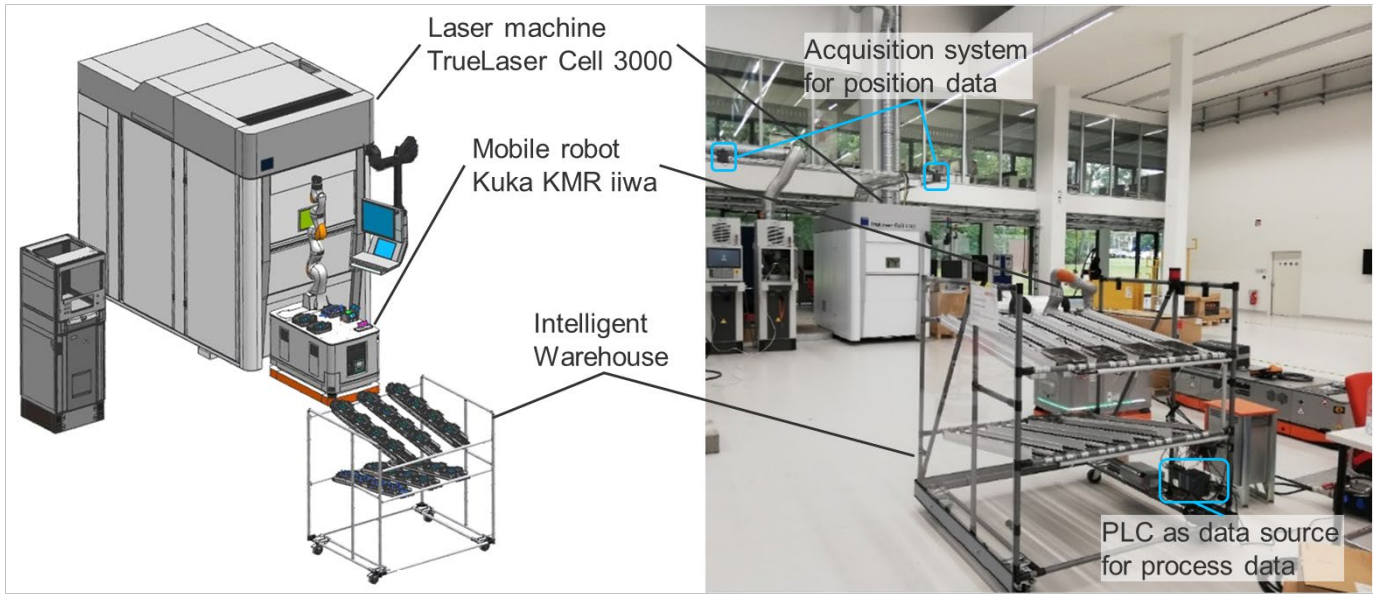

Fig. 3. Brownfield Testbed in the ARENA2036 (right side) and its mechanical model (left side)

The movable robot collects the metal parts in carriers from the warehouse, assembles the model car and transfers it to the welding machine. The welding machine welds the metal parts. The units do not have to be arranged in a fixed, linear chain because of the mobile robot which acts as a driverless transport vehicle and connects the warehouse with the welding machine. Furthermore, the warehouse is equipped with wheels and can therefore easily be moved. This enables together with the mobile robot a variable and easily modifiable production process, which can be rearranged or extended with new units. Therefore, a real-time locating system is installed to identify the position of the single units and the workpiece carries. The cross-domain models of the Digital Twin were created during the design phase using a product lifecycle management system and is the cyber part of the production system. These product lifecycle management systems are able to integrate and manage numerous tools and their models during design, operation and maintenance. The Digital Twin of the system consists of mechanical models, kinematic definitions, electrical models, process simulations and automation software models. This Digital Twin was used, for example, to virtually verify in advance whether the robot could reach the carriers and the clamping devices in the welding machine. In this research the intelligent warehouse is used to implement and test the modeling approach, especially the data fusion of position and process data is addressed in the following sections.

### A. Acquiring data from the brownfield testbed to create the Digital Twin

The required information about the warehouse is obtained from the PLC project and two additional sources. The first source is the I/O signals received directly from the warehouse PLC. For this purpose, the OPC UA server of the PLC is used and the values are retrieved according to the publish and subscribe concept. An OPC UA Client is used to log the data over time and write it into an Influx database. Fig. 4 illustrates the data acquisition process of the I/O signals and the position data through a real time locating system. The position data is obtained via the real time locating system and the location server installed on an industrial computer. The moving RTLS transponders send a heartbeat signal to the anchors/gateway, which add a timestamp. The location server analyses the incoming messages from the anchors/gateways and calculates the transponder position through the time difference. The calculated positions are then exported structured according to the simple location message protocol (SLMP) and are transmitted via a TCP connection from the Siemens Location Manager of the industrial PC to a personal computer. Besides the I/O signals, the position data are stored in the time series database (InfluxDB).

For the presented aspects of the developed modeling approach, the position data of the workpiece carriers in the warehouse are of particular interest. Therefore, six workpiece carriers are pushed and withdrawn to each of the four storage rows inside the lower level of the intelligent warehouse in various combinations. During this procedure the process data and position data are collected for the following information extraction methods.

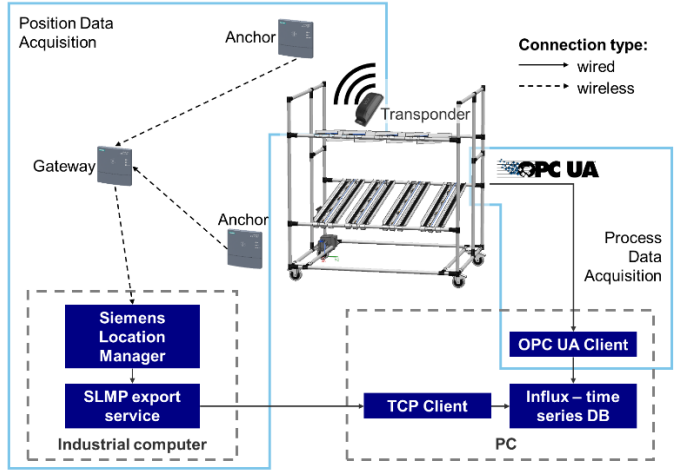

Fig. 4.  Data acquisition from the intelligent warehouse

### B. Data driven analysis to extract the Digital Twin models and relations

The collected position and process data are stored inside time series databases and are analyzed by multiple software modules to extract the contained information about the manufacturing system. Fig. 5 illustrate software modules and the data flow between these modules.

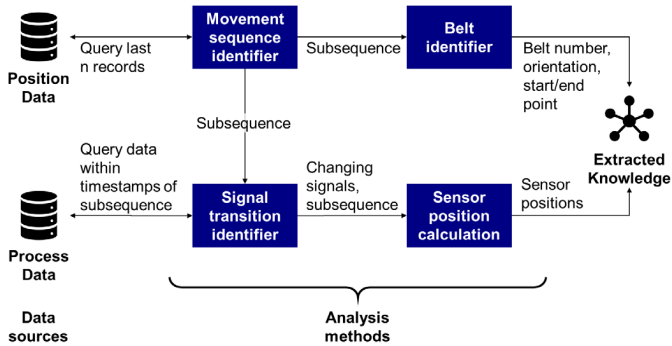

Fig. 5.  Data flow between software modules for data analysis

These software modules are grouped in the following two functionalities:

#### 1) Position data analysis for mechatronic clustering

The position data analysis is done by two modules. The first module is the *movement sequence identifier (see Fig. 5)*, which process and group the position data belonging to the same continuous material movement. Therefor the data is filtered by id of the transponder of the real time locating system, thus the data belonging to one workpiece carrier movement is processed together. Then the position data of a workpiece carrier are analyzed according to continuous movement of the material.

The interruption of the movement of a workpiece carrier is identified by two criteria: (1) a time gap between two data points much longer than the normal update rate, (2) multiple data points containing the same position. The time gap arises due to the transmission rate reduction of the transponders heartbeat signal in rest position. Because minimal shocks are interpreted as movement of the transponders, there are wrong position updates in some cases. Therefore, the second criterion inspects if there are multiple position updates of the same position and thus detects the wrong position updates. With the two-stage separation the subsequences can be identified even with modified real time locating system parameters and continuous position updates. As a result, the movement sequence identifier provides the subsequences consisting of the start-/end-positions, the data points and the timestamps of a workpiece carrier between two stops. These subsequences are passed to the belt identifier and sensor position identification.

The *belt identifier* (see Fig. 5) predicts the storage row the subsequence belongs to. In general, this module predicts the mechatronic function group the position data is affiliated with. The classifier has to solve a multi-class problem to predict the different storage rows of the intelligent warehouse. The input data are the multivariant time series position data from the workpiece carrier movement inside the intelligent warehouse retrieved by the real time locating system and stored in the InfluxDB. The classifier is implemented using a 1-nearest neighbor algorithm with dynamic time warping to compare time sequences with variable length and sampling rate. A prediction accuracy of 86,5% for the data samples is achieved using the one nearest neighbor with dynamic time warping. By comparing the assigned classes to the data samples of one subsequence, determining the majority and readjusting the minorities accordingly, the accuracy is improved.

#### 2) Data fusion analysis extracting the mechanical relations

Besides the separate analysis of the single data source the fused analysis of the position and the process data is prototypical implemented for the intelligent warehouse. By combining the signals and the geometric positions, the approximate positions of the mechatronic automation resources are to be determined. For this purpose, the moving material is assumed to be the trigger for the change of sensor values and its position is used as a reference for the position of the mechatronic components. The sensor position determination is implemented in two modules and requires the two datasets for the position and process data as input. The position data are preprocessed with the module for subsequence identification from Section IV.B.1). The module *signal transition identifier* (see Fig. 5) works based on these subsequences of the workpiece carrier movement inside the warehouse and searches for signal transitions of the light barrier signals of the storage rows in the same time period. If a signal is found that has changed in a subsequence, the signal and the corresponding position of the workpiece carrier at the time of the signal change are passed on together to the sensor position calculation module.

The *sensor position calculation* module (see Fig. 5) extracts the relevant position data samples related to the signal change by clustering all data samples by their position. The largest cluster is formed by the motion data of the responsible workpiece carrier. Random movement data, which is caused by vibrations or simultaneous movements in other storage rows, is categorized in separate clusters. This removes position data of movements that happen to occur at the same time and were mistakenly included from signal transition identifier to the relevant data points. To further improve the position prediction, outliers are removed from the cluster with relevant positions and the remaining positions are averaged.

As a result, the positions of the 16 photoelectric sensors of the lower level of the intelligent warehouse are determined with an accuracy of a few centimeters. The assumption here is that the expansion of the moving material is in the best-case point-shaped and only insignificantly larger than the transponder of the real time locating system. Otherwise, the position of the transponder does not match the sensor trigger point of the material, and thus the responsibility for a signal change. For the workpiece carriers in the warehouse, this assumption could be sufficiently fulfilled, since these are only approximately 15x25 cm in size and the transponder was placed accordingly so that it is centered on the triggering contours of the workpiece carrier.

### C. Resulting data-driven Digital Twin

Using data driven approach analyzing the process and position data of the intelligent warehouse the relation to the mechanical domain, the clusters of assemblies and the dynamic plant behavior is represented. This information combined with the software model from the PLC describes the warehouse and form its Digital Twin.

## V. CONCLUSION AND OUTLOOK

The available methods deliver models of a single engineering domains mechanic, electric or software. Thus, multiple methods need to be combined to reengineer the models from the different dimensions. Especially their relations to the other domains are rarely addressed by them, but are important for the Digital Twin and the system (re-)engineering. The presented approach addresses the creation of multi-dimensional connected digital models. Starting from the PLC as primary information source in particular the software is of an existing manufacturing system modeled but also the relations towards the mechanic and electric are considered using a knowledge graph to combine the information. Using the demonstration system, it is shown in a proof-of-concept that the approach, in particular the combination of process data and position data, delivers the information needed to connect the software signals with the mechanical domain. In particular it is shown that the data driven approach provides the

- relation of the software to the mechanic model,
- plant behavior and
- clustering as mechatronic assembly.

Using this approach, the reengineering of the manufacturing plant as digital models is supported and (partly) automated by the automated connection of engineering domains. Other methods or engineers can enrich the models with details from the mechanical or electrical domain as the relation between the domains are maintained. This results in the Digital Twin of a production system. The entire method uses a manufacturer-independent approaches to collect and analyze the PLC, the dynamic signal traces and the position data. Thus, it is transferable to other automated production systems using different PLC or localization systems with little amendments to the vendor specific data collection modules. In the future research the vendor-independent PLC analysis and data driven analysis of the process data will be addressed more detailed to model the logical and timing dependencies in the software domain. In addition, information storage options and knowledge extraction methods will be further investigated to support and (partially) automate the creation of relevant models for the software reconfiguration based on this information. These models can be used to fasten the reconfiguration and reuse of brownfield manufacturing systems.

ACKNOWLEDGMENT

This work was supported by the Deutsche Forschungsgemeinschaft DFG (German Research Foundation) within the Exzellenzinitiative (Excellence Initiative) – GSC 262 and the Landesministerium für Wissenschaft, Forschung und Kunst Baden-Württemberg (Ministry of Science, Research and the Arts of the State of Baden-Wurttemberg) within the Nachhaltigkeitsförderung (sustainability support) of the projects of the Exzellenzinitiative II.

REFERENCES

[1] S. R. Bader, C. Wolff, M. Vössing, and J.-P. Schmidt, "Towards Enabling Cyber-Physical Systems in Brownfield Environments," in *Exploring Service Science: 9th International Conference Proceedings*, Karlsruhe, Germany, 2018, pp. 165–176.

[2] A. Vaughn, P. Fernandes, and J. T. Shields, "Manufacturing system design framework manual," pp. 1–64, 2002. [Online]. Available: http://hdl.handle.net/1721.1/81902

[3] J. Kiefer, S. Allegretti, and T. Breckle, "Quality- and Lifecycle-oriented Production Engineering in Automotive Industry," *Procedia CIRP*, vol. 62, pp. 446–451, 2017, doi: 10.1016/j.procir.2016.06.086.

[4] B. Ashtari Talkhestani *et al.*, "An architecture of an Intelligent Digital Twin in a Cyber-Physical Production System," *at - Automatisierungstechnik*, vol. 67, no. 9, pp. 762–782, 2019, doi: 10.1515/auto-2019-0039.

[5] J. Kiefer, "Mechatronikorientierte Planung automatisierter Fertigungszellen im Bereich Karosserierohbau," Dissertation, Lehrstuhl für Fertigungstechnik/CAM, Universität des Saarlandes, Saarbrücken, 2007.

[6] M. Strube, "Modellgestützte Modernisierungsplanung industrieller Automatisierungslösungen," Dissertation, Institut für Automatisierungstechnik, Helmut Schmidt Universität, Hamburg, 2014.

[7] S. Biffl, D. Gerhard, and A. Lüder, "Introduction to the Multi-Disciplinary Engineering for Cyber-Physical Production Systems," in *Multi-Disciplinary Engineering for Cyber-Physical Production Systems: Data Models and Software Solutions for Handling Complex Engineering Projects*, S. Biffl, A. Lüder, and D. Gerhard, Eds., Cham: Springer International Publishing, 2017, pp. 1–24.

[8] M. Schamp, S. Hoedt, A. Claeys, E. H. Aghezzaf, and J. Cottyn, "Impact of a virtual twin on commissioning time and quality," *IFAC-PapersOnLine*, vol. 51, no. 11, pp. 1047–1052, 2018, doi: 10.1016/j.ifacol.2018.08.469.

[9] N. Arshad, D. Heimbigner, and A. L. Wolf, "Deployment and dynamic reconfiguration planning for distributed software systems," *Software Qual J*, vol. 15, no. 3, pp. 265–281, 2007, doi: 10.1007/s11219-007-9019-2.

[10] T. Müller, N. Jazdi, J.-P. Schmidt, and M. Weyrich, "Cyber-physical production systems: enhancement with a self-organized reconfiguration management," *Procedia CIRP*, vol. 99, pp. 549–554, 2021, doi: 10.1016/j.procir.2021.03.075.

[11] B. Ashtari Talkhestani, D. Braun, W. Schloegl, and M. Weyrich, "Qualitative and quantitative evaluation of reconfiguring an automation system using Digital Twin," *Procedia CIRP*, vol. 93, pp. 268–273, 2020, doi: 10.1016/j.procir.2020.03.014.

[12] B. Ashtari Talkhestani, "Methodik zur Synchronisierung der Modelle des Digitalen Zwillings automatisierter Systeme," Dissertation, Institut für Automatisierungstechnik und Softwaresysteme, Universität Stuttgart, Stuttgart, 2020.

[13] G. Erdős, T. Nakano, G. Horváth, Y. Nonaka, and J. Váncza, "Recognition of complex engineering objects from large-scale point clouds," *CIRP Annals*, vol. 64, no. 1, pp. 165–168, 2015, doi: 10.1016/j.cirp.2015.04.026.

[14] J. Berglund, E. Lindskog, J. Vallhagen, Z. Wang, C. Berlin, and B. Johansson, "Production System Geometry Assurance Using 3D Imaging," *Procedia CIRP*, vol. 44, pp. 132–137, 2016, doi: 10.1016/j.procir.2016.02.138.

[15] F. Buonamici, M. Carfagni, R. Furferi, L. Governi, A. Lapini, and Y. Volpe, "Reverse engineering modeling methods and tools: a survey," *Computer-Aided Design and Applications*, vol. 15, no. 3, pp. 443–464, 2018, doi: 10.1080/16864360.2017.1397894.

[16] D. Braun, F. Biesinger, N. Jazdi, and M. Weyrich, "A concept for the automated layout generation of an existing production line within the digital twin," *Procedia CIRP*, vol. 97, pp. 302–307, 2020, doi: 10.1016/j.procir.2020.05.242.

[17] G. Koltun, F. Maurer, A. Knoll, E. Trunzer, and B. Vogel-Heuser, "Information Retrieval from Redlined Circuit Diagrams and its Model-Based Representation for Automated Engineering," in *IECON 2018 - 44th Annual Conference of the IEEE Industrial Electronics Society*, Washington D.C., USA, 2018, pp. 3114–3119.

[18] H. Zipper, F. Auris, A. Strahilov, and M. Paul, "Keeping the digital twin up-to-date — Process monitoring to identify changes in a plant," in *Proceedings 2018 IEEE International Conference on Industrial Technology (ICIT)*, Lyon, 2018, pp. 1592–1597.